\documentclass[aps,prd,twocolumn,nofootinbib]{revtex4}

\usepackage{graphicx}

\begin{document}

\title{Scale-Setting Using the Extended Renormalization Group and the Principle of Maximum Conformality: the QCD Coupling Constant at Four Loops}

\author{Stanley J. Brodsky$^{1}$}
\email[email:]{sjbth@slac.stanford.edu}

\author{Xing-Gang Wu$^{1,2}$}
\email[email:]{wuxg@cqu.edu.cn}

\address{$^{1}$ SLAC National Accelerator Laboratory, 2575 Sand Hill Road, Menlo Park, CA 94025, USA\\
$^{2}$ Department of Physics, Chongqing University, Chongqing 401331, P.R. China}

\date{\today}

\begin{abstract}

A key problem in making precise perturbative QCD predictions is to set the proper renormalization scale of the running coupling. The extended renormalization group equations, which express the invariance of the physical observables under both the renormalization scale- and scheme- parameter transformations, provide a convenient way for estimating the scale- and scheme- dependence of the physical process. In this paper, we present a solution for the scale-equation of the extended renormalization group equations at the four-loop level. Using the Principle of Maximum Conformality (PMC) / Brodsky-Lepage-Mackenzie (BLM) scale-setting method, all non-conformal $\{\beta_i\}$-terms in the perturbative expansion series can be summed into the running coupling, and the resulting scale-fixed predictions are independent of the renormalization scheme. The PMC/BLM scales can be fixed order-by-order. As a useful reference, we present a systematic and scheme-independent procedure for setting PMC/BLM scales up to NNLO. An explicit application for determining the scale setting of $R_{e^{+}e^-}(Q)$ up to four loops is presented. By using the world average $\alpha^{\overline{MS}}_s(M_Z) =0.1184 \pm 0.0007$, we obtain the asymptotic scale for the 't Hooft scheme associated with the $\overline{MS}$ scheme, $\Lambda^{'tH}_{\overline{MS}}= 245^{+9}_{-10}$ MeV, and the asymptotic scale for the conventional $\overline{MS}$ scheme, $\Lambda_{\overline{MS}}= 213^{+19}_{-8}$ MeV. \\

\begin{description}
\item[PACS numbers] 12.38.Aw, 11.10.GH, 11.15.Bt
%\item[Keywords] PMC/BLM Scale Setting, Renormalization Scale, Extended Renormalization Group
\end{description}

\end{abstract}

\maketitle

\section{Introduction}

All physical predictions in QCD should in principle be invariant under any choice of renormalization scale and scheme. However at any finite order, the use of different scales and schemes may lead to different theoretical predictions. The optimal procedure for obtaining precise QCD predictions is to choose the renormalization scale so that the result is scheme independent at any fixed order in $\alpha_s$. The result of a scale-setting strategy should satisfy several self-consistent conditions: the existence and uniqueness of the scale, reflexivity, symmetry and transitivity \cite{selfconsistence}. Moreover, perturbative QCD becomes an Abelian theory as $N_c \to 0$, so QCD scale-setting must agree with that of QED in this limit \cite{qedlimit}. We shall show that the Brodsky-Lepage-Mackenzie method (BLM) \cite{blm} and the Principle of Maximum Conformality (PMC) \cite{pmc} provide a solution to this problem \footnote{PMC provides the principle underlying BLM scale setting, so if not specially stated, we usually treat them on equal footing.}.

The main idea of PMC/BLM is that after proper procedures, all non-conformal $\{\beta_i\}$-terms in the perturbative expansion are summed into the running coupling so that the remaining terms in the perturbative series are identical to that of a conformal theory, i.e., the corresponding theory with $\{\beta_i\}=\{0\}$. The QCD predictions from PMC/BLM are then independent of renormalization scheme. It has been found that PMC/BLM satisfies all self-consistent conditions \cite{selfconsistence}. After PMC/BLM scale setting, the divergent ``renormalon" series $(n!\beta_i^{n}\alpha_s^n)$ does not appear in the conformal series; thus as in QED, the scale can be unambiguously set by PMC/BLM.

One can use PMC/BLM to relate perturbative calculable observables in QCD, i.e. to derive commensurate scale relations among different observables \cite{scale1,scale2}. Moreover, from the requirement of scheme- independence, one can determine the displacements among the PMC/BLM scales that are derived under different schemes or conventions. We shall show how to fix the PMC/BLM scales order-by-order. The method for  setting the leading-order (LO) and the next-to-leading order (NLO) PMC/BLM scales has been suggested in the literature \cite{blm,pmc,scale1}. In view of the recent improvements on perturbative QCD loop-calculations and the need to improve the theoretical predictions to confront more accurate experimental data, it is important to provide a systematic and scheme-independent treatment of PMC/BLM up to next-to-next-to-leading order (NNLO).

We shall utilize a generalization of the conventional renormalization group (RG) analysis -- extended RG equations which express the invariance of physical observables under both the renormalization scale- and scheme- parameter transformations \cite{pms,bl1}. In this approach, a universal coupling function which covers all possible choices of scale and scheme is introduced, whose corresponding perturbative series serves as an intermediate device for identifying the scale- and scheme- parameters. It can be treated as a transparent solution to the scale-scheme ambiguity problem. A useful advantage is that the scheme dependence can be reliably estimated through the scheme equations. This approach also provides a platform for a reliable scheme-error analysis and gives a precise definition for the asymptotic scale under a possible renormalization scheme $R$, i.e., the scale for the 't Hooft scheme associated with the $R$-scheme $\Lambda^{'tH}_{R}$ \cite{bl1}. We shall present a general solution for the extended RG equation and give relations between the universal coupling function and the conventional adopted coupling function.

The remaining parts of the paper are organized as follows: in Sec.II, we give the extended RG equations and provide their solution up to four loops. In Sec.III, we present a systematic procedure for setting the PMC/BLM scales up to NNLO. Discussions and an explicit application are also presented in Sec.III. Sec.IV provides a summary.

\section{Extended renormalization group equations}

Conventionally, the scale dependence of an ordinary coupling constant is controlled by the RG equation or the $\beta$-function
\begin{equation} \label{basic-RG}
\beta^R(\alpha^{R}_s)=\frac{d}{d\ln\mu^2}\left(\frac{\alpha^{R}_s(\mu)}{4\pi}\right) =-\sum_{i=0}^{\infty}\beta^{R}_{i}\left(\frac{\alpha^{R}_s(\mu)}{4\pi}\right)^{i+2} ,
\end{equation}
where $R$ stands for an arbitrary renormalization scheme. Various terms in $\beta^R_0$, $\beta^R_1$, $\cdots$, correspond to one-loop and two-loop $\cdots$ contributions respectively. In general, the $\{\beta^R_i\}$ are scheme-dependent and depend on the quark mass ($m_f$) through the variable $m_f^{2}/\mu^2$. According to the decoupling theorem, the quark with mass $m_{f}\gg\mu$ can be ignored, and we can often neglect $m_f$-terms when $m_{f}\ll\mu$. Then, for every renormalization scale $\mu$, we can divide the quarks into active ones with $m_f =0$ and inactive ones that can be ignored. Within this framework, it is well-known that the first two coefficients $\beta^R_{0,1}$ are universal, i.e., $\beta^R_0 \equiv 11-2n_f/3$ and $\beta^R_1 \equiv 102-38n_f/3$ for $n_f$ active flavors. Hereafter, we simply write them as $\beta_0$ and $\beta_1$. Under the $\overline{MS}$-scheme, $\{\beta^{\overline{MS}}_i\}_{i\geq2}$ up to four loops can be found in the literature \cite{beta}.

It will be convenient to use the first two universal coefficients $\beta_0$ and $\beta_1$ to rescale the coupling constant and the scale-parameter $\ln\mu^2$ in Eq.(\ref{basic-RG}). That is, by rescaling the coupling constant as
\begin{displaymath}
a^R=\frac{\beta_1}{4\pi\beta_0}\alpha^R_s
\end{displaymath}
and the scale parameter as
\begin{displaymath}
\tau=\frac{\beta^2_0}{\beta_1} \ln\mu^2 ,
\end{displaymath}
one can express the RG equation (\ref{basic-RG}) into the following simpler canonical form
\begin{equation}\label{scale0}
\frac{d a^R}{d \tau} = -(a^R)^2 \left[1+ a^R +c^R_2 (a^R)^2+c^R_3 (a^R)^3 +\cdots \right] ,
\end{equation}
where $c^R_i = {\beta^R_i \beta_0^{i-1}} / {\beta^i_1}$ for $i=2, 3, \cdots$.

As an extension of the ordinary coupling constant, one can define a universal coupling constant $a(\tau,\{c_i\})$ to include the dependence on the scheme parameters $\{c_i\}$, which satisfies the following extended RG equations \cite{pms,bl1}
\begin{equation}
\beta(a,\{c_i\}) = \frac{\partial a}{\partial \tau} = -a^2 \left[1+ a +c_2 a^2+c_3 a^3 +\cdots \right] \label{scale}
\end{equation}
and
\begin{equation}
\beta_n(a,\{c_i\}) = \frac{\partial a}{\partial c_n} = -\beta(a,\{c_i\}) \int_0^{a} \frac{ x^{n+2} dx}{\beta^2(x,\{c_i\})}  \label{scheme}
\end{equation}
The scale-equation (\ref{scale}), similar to Eq.(\ref{scale0}), can be used to evolve the universal coupling function from one scale to another. By comparing Eq.(\ref{scale0}) with Eq.(\ref{scale}), there exists a value of $\tau=\tau_R$ for which
\begin{equation}\label{asrelation}
a^R(\tau_R)=a(\tau_R,\{c^R_i\}) .
\end{equation}
This shows that any coupling constant $a^R(\tau)$ can be expressed by a universal coupling constant $a(\tau,\{c_i\})$ under the proper correspondence. The scheme-equation (\ref{scheme}) can be used to relate the coupling functions under different schemes by changing $\{c_i\}$. It is noted that the universal coupling function has a particularly simple form when all the scheme parameters $\{c_i\}$ are set to zero, i.e., the coupling function can be written as a function of the scale in terms of the Lambert $W$ function \cite{lambert}. This special case with $\{c_i\}\equiv \{0\}$ is usually called the 't Hooft scheme \cite{tH}. In addition to simplifying the solution of the RG equations, the 't Hooft scheme also provides a precise definition for the asymptotic scale ($\Lambda$) of QCD as will be shown below \footnote{Recently, it has been found that the 't Hooft scheme fails to reproduce the factorized form of the $\overline{MS}$-scheme generalization of the generalized Crewther relation \cite{kataev}. This shows that one cannot use it for studying some special theoretical features of gauge theories beyond the two-loop level. Additional references and detailed discussions of the complimentary
approach by Kataev et al. may be found in Ref.~\cite{kataev}}.

The evolution of the universal running coupling can be obtained by integrating Eq.(\ref{scale}), which can be rewritten as
\begin{equation}
\left(\frac{\beta^2_0}{\beta_1}\ln\frac{\mu^2}{\mu^2_0}\right) = \int^{a(\tau,\{c_i\})}_{a(\tau_0,\{c_i\})}\frac{d a}{\beta(a,\{c_i\})} ,
\end{equation}
where $\tau_0=({\beta^2_0}/{\beta_1}) \ln\mu_0^2$ with $\mu_0$ stands for an initial scale. Up to ${\cal O}(a^3)$, it leads to
\begin{equation}
L ={\cal C} +\frac{1}{a} + \ln a +\left(c_2-1 \right) a + \frac{c_3-2 c_2+1}{2}a^2 + {\cal O}(a^3), \label{scale2}
\end{equation}
where $L =({\beta^2_0}/{\beta_1})\ln(\mu^2/\Lambda^2)$ and ${\cal C}$ is an arbitrary integration constant. $\Lambda$ stands for the asymptotic scale, which is scale-invariant and leads to the coupling constant without any reference to the specific initial scale $\mu_0$ \cite{conLam1,conLam2}. The value of $\Lambda$ can be associated with the typical hadron size, which is not predicted by the QCD theory but must be extracted from a measurement of strong coupling constant at a given reference scale or a QCD measure with mass dimensions such as the pion decay constant $f_\pi$. The values of ${\cal C}$ and $\Lambda$ are correlated with each other. One can find a general relation between the asymptotic parameters under different schemes from Eq.(\ref{scale2}), i.e. for their values under two different schemes $R$ and $S$, we have
\begin{equation}
\exp\left[\frac{\beta_1}{2\beta_0^2} {\cal C}_S \right]\Lambda_S =\exp\left[\frac{\beta_1}{2\beta_0^2} {\cal C}_R \right] \Lambda_R , \label{general}
\end{equation}
where $\Lambda_S$ and $\Lambda_R$ are asymptotic scales corresponding to the choice of the integration constants ${\cal C}_S$ and ${\cal C}_R$ respectively.

The 't Hooft scheme is free of higher-order corrections, i.e., all higher-order coefficients $\{c_i\}\equiv \{0\}$, and then its coupling constant is given by the solution of
\begin{equation}
L^{'tH}= \frac{1}{a^{'tH}} + \ln\left(\frac{a^{'tH}}{1+a^{'tH}}\right) ,
\end{equation}
where $L^{'tH} =({\beta^2_0}/{\beta_1})\ln(\mu^2/\Lambda^{'tH2})$ and the integration constant ${\cal C}$ has been absorbed into the asymptotic scale $\Lambda^{'tH}$ for convenience. The 't Hooft coupling constant presents a formal singularity at $L^{'tH}=0$; i.e. $a^{'tH}\equiv a(0,\{0\})=\infty$. Inversely, it can provide a precise definition for the asymptotic scale; i.e., the 't Hooft scale $\Lambda^{'tH}$, which is defined to be the pole of the coupling function in the 't Hooft scheme, $a^{'tH}\equiv a({\beta^2_0}/{\beta_1} \ln(\mu^2/\Lambda^{'tH2}),\{0\})$. Since the absorbed integration constant ${\cal C}$ is arbitrary, the value of $\Lambda^{'tH}$ is not unique, and there are infinite number of 't Hooft schemes, differing only by the value of $\Lambda^{'tH}$. However, under a specific renormalization scheme ($R$-scheme), its asymptotic scale can be fixed to be the 't Hooft scale associated with the $R$-scheme $\Lambda^{'tH}_R$ \cite{bl1}, which enters into both $a^{R}(\mu)= a({\beta^2_0}/{\beta_1} \ln(\mu^2/\Lambda^{'tH2}_R),\{c^R_i\})$ and $a^{'tH}(\mu)= a({\beta^2_0}/{\beta_1} \ln(\mu^2/\Lambda^{'tH2}_R),\{0\})$. Here the word ``associated" means we are choosing the particular 't Hooft scheme that shares the same 't Hooft scale with the $R$-scheme. In practice, one can obtain a relation between $\Lambda^{'tH}_R$ and the asymptotic scale $\Lambda_R$ for $R$-scheme by setting ${\cal C}_S=0$ (here $S$ stands for the 't Hooft scheme) in the left-hand side of Eq.(\ref{general}), i.e.
\begin{equation}
\Lambda^{'tH}_R=\exp\left(\frac{\beta_1}{2\beta_0^2}{\cal C}_R\right) \Lambda_R .
\end{equation}
As a special case, by choosing ${\cal C}_{\overline{MS}}=\ln{\beta_0^2}/{\beta_1}$ \cite{conLam1,conLam2}, we obtain
\begin{equation}\label{relation}
\Lambda^{'tH}_{\overline{MS}}=\left(\frac{\beta_1}{\beta_0^2}\right)^{-\beta_1/2\beta_0^2} \Lambda_{\overline{MS}} ,
\end{equation}
which agrees with the observation presented in Ref.\cite{bl1}. The present definition of $\Lambda_{\overline{MS}}$ associated with the choice of ${\cal C}_{\overline{MS}}=\ln{\beta_0^2}/{\beta_1}$ is the conventional one, originally suggested in Refs.\cite{conLam1,conLam2}; there are other choices for ${\cal C}_{\overline{MS}}$ \cite{lams1,lams2,lams3}, which can be helpful in certain cases.

Eq.(\ref{scale2}) may be solved iteratively, and the solution to the universal coupling constant can be expanded as a power series of $1/L$; i.e., up to four loops,
\begin{widetext}
\begin{eqnarray}\label{alphas}
a &=& \frac{1}{L}+ \frac{1}{L^2}\left({\cal C}- \ln L\right) + \frac{1}{L^3}\left[{\cal C}^2 +{\cal C} +c_2 -(2{\cal C}-\ln L +1)\ln L -1\right] + \nonumber\\
&&\frac{1}{L^4}\left\{ {\cal C}\left({\cal C}^2 +\frac{5}{2}{\cal C} + 3 c_2 -2\right) -\frac{1-c_3}{2} -\left[3{\cal C}^2 +5{\cal C} +3c_2-2 -\left(3{\cal C} -\ln L +\frac{5}{2}\right)\ln L \right]\ln L\right\} +{\cal O}\left(\frac{1}{L^5}\right) .
\end{eqnarray}
\end{widetext}
As a cross-check, one finds that the above solution agrees with Ref.\cite{fourloopa} after proper parameter transformations and by identifying the integration constant ${\cal C}^*$ used there to be ${\cal C}^* = \frac{\beta_1}{\beta_0^2}\left({\cal C}-\ln\frac{4\beta_0}{\beta_1}\right)$. When setting $\{c_i\}=\{0\}$ and ${\cal C}=0$, we recover the coupling constant under the 't Hooft scheme.

\section{BLM scale-setting up to NNLO}

Generally, perturbative QCD prediction for a physical observable $\rho$ can be written as
\begin{eqnarray}
\rho &=& r_0 \Big[ a^n_s(Q) +(A_{1}+A_{2} n_f) a^{n+1}_s(Q) \nonumber\\
&&+ (B_{1} +B_{2}n_f + B_{3}n_f^2 ) a^{n+2}_s(Q) \nonumber\\
&& + (C_{1} +C_{2}n_f + C_{3}n_f^2 + C_{4} n_f^3 ) a^{n+3}_s(Q)+ \cdots \Big] \label{eq_blm}
\end{eqnarray}
where $a_s(Q)={\alpha_s(Q)}/{\pi}$ and the overall tree-level parameter $r_0$ is scale-independent and is free of $a_s(Q)$. Here $n_f$ stands for the quark flavor number and $n (\geq1)$ stands for the initial $\alpha_s$-order at the tree level. After proper scale-setting, all $n_f$-terms in the perturbative expansion can be summed into the running coupling. Here, we shall concentrate on those processes in which all $n_f$-terms are associated with the $\{\beta_i\}$-terms. Note that in higher-order processes, there are $n_f$-terms coming from the Feynman diagrams with the light-by-light quark loops which are irrelevant to the ultra-violet cutoff. Those $n_f$-terms have no relation to $\{\beta_i\}$-terms \cite{blm}, and they should be identified and kept separate from the BLM scale setting \footnote{Those $n_f$-terms, coming from the light-quark loops connected to at least four photon/gluon lines, are of higher twist and are usually power suppressed by the hard scales, so they can be safely neglected in typical applications.}.

The BLM scales can be determined in a general scheme-independent way. The generalization of the BLM procedure to higher order assigns a different renormalization scale for each order in the perturbative series. We can shift the renormalization scale $Q$ into effective ones until we fully absorb those higher-order terms with $n_f$-dependence into the running coupling \footnote{Another way to set the BLM scale up to NNLO can be found in Refs.\cite{Kataev,Mikhailov1}, where however a unified effective scale $Q^*$ is adopted for all orders.}. LO and NLO BLM scale setting have been done in the literature \cite{blm,scale1}. Because of recent improvements in perturbative QCD loop-calculations, it is important  to provide a systematic and scheme-independent treatment of BLM up to NNLO. The BLM scales can be fixed order by order. In the following, we show how to set the BLM scales for the observable $\rho$.

More explicitly, the first step of the BLM method is to set the effective scale $Q^{*}$ at LO
\begin{eqnarray}
\rho &=& r_0\Big[ a^n_s(Q^*) + \widetilde{A}_{1} a^{n+1}_s(Q^*) + (\widetilde{B}_{1} + \widetilde{B}_{2}n_f) a^{n+2}_s(Q^*) \nonumber\\
&& \quad\quad + (\widetilde{C}_{1} +\widetilde{C}_{2}n_f +\widetilde{C}_{3}n_f^2) a^{n+3}_s(Q^*)+\cdots \Big]\ . \label{first}
\end{eqnarray}
The second step is to set the effective scale $Q^{**}$ at NLO
\begin{eqnarray}
\rho &=& r_0\Big[ a^n_s(Q^*) + \widetilde{A}_{1} a^{n+1}_s(Q^{**}) + \widetilde{\widetilde{B}}_{1} a^{n+2}_s(Q^{**}) \nonumber\\
&& \quad\quad + (\widetilde{\widetilde{C}}_{1} +\widetilde{\widetilde{C}}_{2}n_f) a^{n+3}_s(Q^{**})+\cdots \Big]\ , \label{second}
\end{eqnarray}
and the final step is to set the effective scale $Q^{***}$ at NNLO
\begin{eqnarray}
\rho &=& r_0\Big[ a^n_s(Q^*) + \widetilde{A}_{1} a^{n+1}_s(Q^{**}) + \widetilde{\widetilde{B}}_{1} a^{n+2}_s(Q^{***}) \nonumber\\
&& \quad\quad +\widetilde{\widetilde{\widetilde{C}}}_{1} a^{n+3}_s(Q^{***})+\cdots \Big]\ . \label{third}
\end{eqnarray}
The step-by-step coefficients are presented in the Appendix. When performing the scale shifts $Q\to Q^*$, $Q^* \to Q^{**}$ and $Q^{**}\to Q^{***}$, we eliminate the $n_f$-terms associated with the $\{\beta_i\}$-terms completely. At the same time, we also have to modify the coefficients, since the net changes to the coefficients are proportional to $\beta$-functions. To set the effective scale for $a^{n+3}_s$, one needs even higher order information and here, a sensible choice is $Q^{***}$, since this is the renormalization scale after shifting the scales up to NNLO. Note that the effective scales should be a perturbative series of $a_s$ so as to absorb all $n_f$-dependent terms properly, and up to NNLO, three effective scales can be written as
\begin{eqnarray}
&&\ln\frac{Q^{*2}}{Q^2} = \ln \frac{Q^{*2}_0}{Q^2} + \frac{x \beta_0}{4} \ln \frac{Q^{*2}_0}{Q^2} a_s(Q) +\frac{y}{16}\left(\beta^2_0 \ln^2 \frac{Q^{*2}_0}{Q^2}\right.\nonumber\\
&&\quad\quad\quad\quad \left. -\beta_1 \ln\frac{Q^{*2}_0}{Q^2}\right)a^2_s(Q) + {\cal O}(a^3_s)\\
&&\ln\frac{Q^{**2}}{Q^{*2}} = \ln\frac{Q^{**2}_0}{Q^{*2}} + \frac{z \beta_0}{4} \ln \frac{Q^{**2}_0}{Q^{*2}} a_s(Q^*)+ {\cal O}(a^2_s)  \\
&&\ln\frac{Q^{***2}}{Q^{**2}} = \ln \frac{Q^{***2}_0}{Q^{**2}}+ {\cal O}(a_s)
\end{eqnarray}
where the effective scales $Q_0^{*,**,***}$ are determined so as to eliminate $A_2 n_f$, $\widetilde{B}_2 n_f$ and $\widetilde{\widetilde{C}}_2 n_f$-terms completely, the parameters $x$ and $z$ are used to eliminate the $B_3 n_f^2$ and the $\widetilde{C}_3 n_f^2$ terms respectively, and the parameter $y$ is used to eliminate the $C_4 n_f^3$-term. It is found that
\begin{eqnarray}
\ln \frac{Q^{*2}_0}{Q^2} &=& \frac{6A_2}{n}\\
\ln \frac{Q_0^{**2}}{Q^{*2}} &=& \frac{6\widetilde{B}_2}{(n+1)\widetilde{A}_1} \\
\ln \frac{Q_0^{***2}}{Q^{**2}} &=& \frac{6\widetilde{\widetilde{C}}_2}{(n+2)\widetilde{\widetilde{B}}_1}
\end{eqnarray}
and
\begin{eqnarray}
x &=& \frac{3(n+1)A_2^2 -6 n B_3}{n A_2} \\
y &=& \frac{(n+1)(2n+1)A_2^3 -6n(n+1)A_2 B_3 +6n^2 C_4}{n A^2_2} \\
z &=& \frac{3(n+2)\widetilde{B}_2^2 -6(n+1)\widetilde{A}_1 \widetilde{C}_3}{(n+1) \widetilde{A}_1 \widetilde{B_2}}
\end{eqnarray}
The coefficients $A_{i}$, $B_{i}$, $C_{i}$ and etc. are renormalization-scheme dependent, so different renormalization schemes lead to different BLM scales $Q^{*,**,***}$; however the final result for $\rho$ should be scheme independent. Using this argument, one can use BLM scale-setting method to relate perturbatively calculable observables; i.e. to derive commensurate scale relations among different observables \cite{scale1}. In fact, any perturbatively-calculable physical observable can be used to define an effective coupling constant by incorporating the entire radiative correction into its definition \cite{eff}. For example $R_{e^+e^-}(Q) \equiv R^0_{e^+e^-}(Q)\left[1+{\alpha^R_s(Q)\over \pi}\right]$ defines an effective coupling constant $\alpha^R_s(Q)$, where $R^0_{e^+ e^-}(Q)$ stands for the Born result. Any effective coupling constant can be used as a reference coupling constant to define the renormalization procedure. More generally, each effective running coupling constant or renormalization scheme is a special case of the universal coupling function as shown by Eq.(\ref{asrelation}).

The NLO commensurate scale relations between different effective coupling constants can be found in Ref.\cite{scale1}. Replacing the observable $\rho$ by its corresponding effective coupling constant and changing $a_s$ to be another effective coupling constant, starting from Eq.(\ref{eq_blm}) and following the same procedures, one can naturally obtain the commensurate scale relations up to NNLO. Moreover, by using the relations between $Q^{*,**,***}$ and $Q$, one can find the needed scale displacement among the effective scales which are derived under different schemes or conventions so as to ensure the scheme-independence of the observables. For example, from the relation between $Q^{*}$ and $Q$, one can easily obtain the well-known one-loop relation for the coupling constant \cite{blm}, $\alpha_s^{\overline{MS}}(e^{-5/3}Q^2) =\alpha_s^{GM-L}(Q^2)$, where the scale displacement $e^{-5/3}$ between the $\overline{MS}$ scheme and the Gell-Mann-Low scheme \cite{gml} is a result of the convention that is chosen to define the minimal dimensional regularization scheme \cite{conLam1}.

\subsection{The PMC and BLM correspondence principle}

A systematic procedure for setting PMC scale at LO has been suggested in Ref.\cite{pmc}. The main procedure is to distinguish the nonconformal terms from the conformal terms by the variation of the cross section with respect to $\ln\mu_0^2$ ($\mu_0$ stands for some initial scale of the process). At LO, there is only one type of $\{\beta_i\}$-term (i.e. $\beta_0$) and the nonconformal terms always have the form of $\beta_0 \ln\mu_0^2$, so one can determine the nonconformal terms exactly. However, at higher orders, the $\ln\mu_0^2$-terms usually appear in a power series as $\beta_0 \ln\mu_0^2$, $\beta_1 \ln\mu_0^2$, $\beta_0^2 (\ln\mu_0^2)^2$, etc.. So this method is no longer adaptable to deal with the higher-order corrections, because the derivative with respect to a single $\ln\mu_0^2$ cannot distinguish all the emerged $\{\beta_i\}$-terms. Some alternative should be introduced.

The purpose of the running coupling in any gauge theory is to sum up all the terms involving the $\{\beta_i\}$-functions, conversely, one can find all the needed $\{\beta_i\}$-terms at any relevant order from the expansion of the running coupling in the form of Eq.(\ref{alphaexp}). Using this fact and also the known relation between $\{\beta_i\}$ and $n_f$, one can obtain the PMC scales from the BLM scale-setting method. We call this the PMC and BLM correspondence principle. Since $\{\beta_i\}$ ($i\geq2$) are scheme-dependent, the PMC and BLM correspondence depends on the renormalization scheme beyond the two-loop level. It is noted that such an expansion is different from that of Refs.\cite{Mikhailov1,Mikhailov2}, where all $\{\beta_i\}$-terms which may contribute at the same order have been introduced to deal with the Adler $D$-function.

More explicitly, up to NNLO, the physical observable $\rho$ can be expanded in $\{\beta_i\}$-series as,
\begin{eqnarray}
\rho &=& r_0 \Big[a^n_s(Q) +(A^0_{1}+ A^0_{2} \beta_0) a^{n+1}_s(Q) \nonumber\\
&& +(B^0_{1}+ B^0_{2} \beta_1 + B^0_{3} \beta_0^2) a^{n+2}_s(Q) \nonumber\\
&& +(C^0_{1}+ C^0_{2} \beta_2 + C^0_{3} \beta_0 \beta_1 + C^0_{4}\beta_0^3) a^{n+3}_s(Q)\Big] . \label{eq_pmc}
\end{eqnarray}
The results for PMC can be naturally obtained from the BLM scale setting through proper parameter correspondence, i.e.
\begin{eqnarray}
A_{1} &=& A^0_{1}+11 A^0_{2} \\
A_{2} &=& -\frac{2}{3}A^0_{2} \\
B_{1} &=& B^0_{1}+102 B^0_{2}+121 B^0_{3} \\
B_{2} &=& -\frac{2}{3}(19B^0_{2}+22B^0_{3}) \\
B_{3} &=& \frac{4}{9}B^0_{3} \\
C_{1} &=& C^0_{1}+\frac{2857}{2}C^0_{2}+1122C^0_{3}+1331C^0_{4} \\
C_{2} &=& -\frac{1}{18}(5033C^0_{2}-3732C^0_{3}-4356C^0_{4}) \\
C_{3} &=& \frac{1}{54}(325C^0_{2}+456C^0_{3}+792C^0_{4}) \\
C_{4} &=& -\frac{8}{27}C^0_{4}
\end{eqnarray}
which are obtained with the help of Eqs.(\ref{eq_blm},\ref{eq_pmc}) and the four-loop $\{\beta_i\}$-terms  under the $\overline{MS}$ scheme \cite{beta}.

\subsection{An application of PMC/BLM scale setting}

We present an application of PMC/BLM scale setting up to NNLO by dealing with the total hadronic cross section in $e^{+}e^{-}$ annihilation, $R_{e^{+}e^{-}}(Q)=R(e^{+}e^{-}\to{\rm hadrons})$. The explicit expression for $R_{e^{+}e^{-}}(Q)$ up to $\alpha_s^4$-order under the $\overline{MS}$-scheme can be found in Ref.\cite{Ralphas}. One finds
\begin{widetext}
\begin{eqnarray}
R_{e^{+}e^{-}}(Q) = 3\sum_q e_q^2 \Bigg[ 1  + \left(a^{\overline{MS}}(Q)\right) +  A \left(a^{\overline{MS}}(Q)\right)^2 + B \left(a^{\overline{MS}}(Q)\right)^3 + C \left(a^{\overline{MS}}(Q)\right)^4 \Bigg] ,
\end{eqnarray}
\end{widetext}
where
\begin{eqnarray}
A &=& 1.9857 - 0.1152 n_f , \nonumber\\
B &=& -6.63694 - 1.20013 n_f - 0.00518 n_f^2 -1.240 \eta , \nonumber\\
C &=& -156.61 + 18.77 n_f - 0.7974 n_f^2  + 0.0215 n_f^3 +\kappa \eta , \nonumber
\end{eqnarray}
where $\eta={\left(\sum_q e_q\right)^2}/{\left(3\sum_q e_q^2\right)}$, $e_q$ is the electric charge for the active flavors. The coefficient $\kappa$ is yet to be determined, its value is small \cite{fourloopa,fourloopb,kataevC,Chetyrkin} and its contribution will be further suppressed by the factor $\eta$, so we have set its value to zero in the following numerical calculation. The values of $A$, $B$ and $C$ for $n_f=3$, $4$ and $5$ are presented in Tab.\ref{tab}.

\begin{widetext}
\begin{center}
\begin{table}
\begin{center}
\caption{Coefficients for the perturbative expansion of $R_{e^{+}e^{-}}(Q)$ before and after  BLM scale-setting. }
\begin{tabular}{|c||c|c|c|}
\hline\hline
   ~~~ ~~~   & ~~~$n_f =3$~~~ & ~~~$n_f =4$~~~ & ~~~$n_f =5$~~~ \\
\hline\hline
$A$ & 1.6401  &  1.5249 &  1.4097 \\
\hline
$B$ & -10.2840  &  -11.6857 & -12.8047 \\
\hline
$C$ & -106.8960  & -92.9124 +$2\kappa/15$   & -80.0075 +$\kappa/33$  \\
\hline\hline
$\widetilde{A}$ &  0.0849  &  0.0849  &  0.0849  \\
\hline
$\widetilde{\widetilde{B}}$ & -23.2269  &  -23.3923  &  -23.2645  \\
\hline
$\widetilde{\widetilde{\widetilde{C}}}$ &  82.3440  &  82.3440 +$2\kappa/15$  & 82.3440 +$\kappa/33$  \\
\hline\hline
\end{tabular}
\label{tab}
\end{center}
\end{table}
\end{center}
\end{widetext}

At the present order in $\alpha_s$, the $n_f$-terms which come from the light-by-light quark loops and are irrelevant to the ultra-violet cutoff do not emerge, so all $n_f$-terms in the above equation should be fully absorbed into $\alpha_s$. After doing the BLM scale setting up to NNLO, we obtain
\begin{widetext}
\begin{eqnarray}
R_{e^{+}e^{-}}(Q) = 3\sum_q e_q^2 \Bigg[ 1  + \left(a^{\overline{MS}}_s(Q^*)\right) +  \widetilde{A} \left(a^{\overline{MS}}_s(Q^{**})\right)^2 + \widetilde{\widetilde{B}} \left(a^{\overline{MS}}_s(Q^{***})\right)^3 +\widetilde{\widetilde{\widetilde{C}}} \left(a^{\overline{MS}}_s(Q^{***})\right)^4 \Bigg] ,
\end{eqnarray}
\end{widetext}
where all the coefficients and effective scales can be calculated with the help of the formulae listed in the Appendix. The coefficients are presented in TAB.\ref{tab}, slight differences for $\widetilde{\widetilde{B}}$ and $\widetilde{\widetilde{\widetilde{C}}}$ with varying $n_f$ are caused by the charge-dependent parameter $\eta$.

From the experimental value, $r_{e^+ e^-}(31.6GeV)=\frac{3}{11} R_{e^+ e^-}(31.6GeV)=1.0527\pm0.0050$ \cite{rexp}, we obtain
\begin{equation}
\Lambda^{'tH}_{\overline{MS}} = 412^{+206}_{-161} {\rm MeV}
\end{equation}
and
\begin{equation}
\Lambda_{\overline{MS}}       = 359^{+181}_{-140} {\rm MeV} .
\end{equation}
With the help of the four-loop coupling constant (\ref{alphas}), we obtain $\alpha^{\overline{MS}}_s(M_Z)=0.129^{+0.009}_{-0.010}$. This value is somewhat larger than the present world average $\alpha^{\overline{MS}}_s(M_Z) =0.1184 \pm 0.0007 $ \cite{pdg}. However, it is consistent with the values obtained from $e^+ e^-$ colliders, i.e. $\alpha^{\overline{MS}}_s(M_Z)=0.13\pm 0.005\pm0.03$ by the CLEO Collaboration \cite{cleo} and $\alpha^{\overline{MS}}_s(M_Z)=0.1224\pm 0.0039$ from the jet shape analysis \cite{jet}. One may observe that a smaller central value of the world average for $\alpha^{\overline{MS}}_s(M_Z)$ also results from the measurements of $\tau$-decays, $\Upsilon$-decays, the jet production in the deep-inelastic-scattering processes, and from heavy quarkonia based on the unquenched QCD lattice calculations \cite{aveas}. A larger $\Lambda_{\overline{MS}}$ leads to a larger $\alpha^{\overline{MS}}_s(M_Z)$, and vice versa. If we set $\alpha^{\overline{MS}}_s(M_Z)$ to the present world average, we obtain $\Lambda^{'tH}_{\overline{MS}}|_{n_f=5}= 245^{+9}_{-10}$ MeV and $\Lambda_{\overline{MS}}|_{n_f=5}= 213^{+19}_{-8}$ MeV \footnote{ Ref.\cite{aveas} obtained a slightly different value of $\Lambda_{\overline{MS}}|_{n_f=5}= 215 \pm 9 MeV$. However, it is obtained by taking a wrong sign of $({\beta_3}/{2\beta_0})$ in the four-loop terms, which should be negative rather than positive.}.

It is found that after PMC/BLM scale-setting, the perturbative expansion of $R_{e^+ e^-}(Q=31.6GeV)$ becomes more convergent. The higher-order corrections are used to set the scales $Q^*$, $Q^{**}$ and etc., respectively. In particular, we find $Q^*=\left(0.757\pm0.008\right)Q$ which leads to $a^{\overline{MS}}_s(Q^*) /a^{\overline{MS}}_s(Q) =1.060\pm0.004$.

As a final remark, one can estimate the error caused by $\kappa$ with the help of the scheme-dependent equation (\ref{scheme}). Such an analysis has been done in Ref.\cite{bl1} \footnote{Note there is a typo in Eq.(48) of  Ref.\cite{bl1}, which should be changed to, $a_0={a_+}/{\left(1+\frac{3}{2}c^R_3 a^3_+ \right)^{1/3}}$.}. It is found that even if we set its value that leads to the $\kappa$-term has a comparable magnitude with those without $\kappa$ at the fourth-order, we shall only achieve an additional $2\%$ scheme error in addition to the above experimental errors.

\section{Summary}

The extended renormalization group equations provide a convenient way for estimating both the scale- and scheme- dependence of the QCD predictions for a physical process. The scheme dependence of a process can be reliably estimated by the scheme-equations for the extended renormalization group.

In the present paper, we have presented a general solution to the scale equation of the extended renormalization group equations at the four-loop level. This formalism provides a platform for a reliable error analysis and also provides a precise definition for the asymptotic scale under any renormalization $R$-scheme, $\Lambda^{'tH}_R$, which is defined as the pole of the strong coupling constant in the 't Hooft scheme associated with $R$-scheme.

We have also given a systematic and renormalization scheme-independent method for setting the PMC/BLM scales up to NNLO. The PMC provides the principle underlying BLM scale setting; the two methods are equivalent to each other through the PMC and BLM correspondence principle. The scales can be set unambiguously by PMC/BLM, which allows us to set the renormalization scale at any required orders to obtain a scheme-independent result. Such a scheme-independence can be adopted to derive commensurate scale relations among different observables and to find the displacements among the effective PMC/BLM scales which are derived under different schemes or conventions.

The elimination of the renormalization scale ambiguity and the scheme dependence using PMC/BLM will not only increase the precision of QCD tests, but it will also increase the sensitivity of the collider experiments to new physics beyond the Standard Model.

\hspace{2cm}

{\bf Acknowledgements}: The authors thank Leonardo Di Giustino for helpful discussions. This work was supported in part by the Program for New Century Excellent Talents in University under Grant NO.NCET-10-0882, Natural Science Foundation of China under Grant NO.10805082 and NO.11075225, and the Department of Energy contract DE-AC02-76SF00515.

\appendix

\section{Coefficients for the BLM scale setting up to NNLO}

The step-by-step coefficients for the BLM scale setting up to NNLO, which are introduced in Eqs.(\ref{first},\ref{second},\ref{third}), are listed in the following
\begin{widetext}
\begin{eqnarray}
\widetilde{A}_1 &=& A_1 +\frac{33}{2}A_2 \;,\; \widetilde{\widetilde{B}}_1 = \widetilde{B}_1 +\frac{33}{2}\widetilde{B}_2 \;,\; \widetilde{\widetilde{\widetilde{C}}}_1 = \widetilde{\widetilde{C}}_1 +\frac{33}{2}\widetilde{\widetilde{C}}_2\\
\widetilde{B}_1 &=& \frac{1}{4n}\Bigg[1089(n+1)A_2^2 +153n A_2 +66(n+1)A_1 A_2+(4B_1 -1089 B_3)n\Bigg] \\
\widetilde{B}_2 &=&\frac{-1}{4n}\Bigg[66(n+1)A_2^2 +19n A_2 +4(n+1)A_1 A_2 -4n(B_2 +33B_3)\Bigg] \\
\widetilde{C}_1 &=& \frac{1}{64{A_2}n^2} \Bigg[-40392{C_4}n^3 + 143748{{A_2}}^4(3+5n+2n^2)  + 8{A_2}n^2 ( 8{C_1} + 35937{C_4} + \nonumber\\
&& 5049{B_3}n )  -6732{{A_2}}^3n(2n^2-6n -11)+72{A_1}{A_2}(1+n)(34{A_2}n - 242{B_3}n +\nonumber\\
&& 121{{A_2}}^2( 3 + 2n ))  + 3{{A_2}}^2n( 2857n + 352{B_1}( 2 + n )  -  95832{B_3}( 3 + 2n )  ) \Bigg]\\
\widetilde{C}_2 &=& \frac{1}{192 {A_2}n^2}\Bigg[22392{C_4}n^3 - 52272{{A_2}}^4(3+5n+2n^2 ) - 24{A_2}n^2 ( -8{C_2} + \nonumber\\
&& 6534{C_4} + 933{B_3}n )  - 48{A_1}{A_2}( 1 + n ) ( 19{A_2}n - 132{B_3}n + 66{{A_2}}^2( 3 + 2n )  )  + \nonumber\\
&& {A_2}^2n( -5033n - 192{B_1}( 2 + n ) + 3168{B_2}( 2 + n ) + 52272{B_3}( 8 + 5n ))+ \nonumber\\
&& 12{{A_2}}^3n( -2809 + 2n( -627 + 311n )  ) \Bigg]\\
\widetilde{C}_3 &=& \frac{1}{576{A_2}n^2}\Bigg[ -2736{C_4}n^3 + 4752{{A_2}}^4(3+5n+2n^2)  + 144{A_2}n^2     ( 4{C_3} + 198{C_4} + \nonumber\\
&& 19{B_3}n )  + 456{{A_2}}^3 n (5-2 n^2)+288{A_1}{A_2}(1+n)(-2{B_3}n + {{A_2}}^2(3 + 2n )) \nonumber\\
&&  -{{A_2}}^2n( -325n + 576{B_2}( 2 + n )  + 9504{B_3}( 5 + 3n )  )\Bigg]\\
\widetilde{\widetilde{C}}_1 &=& \frac{1}{4(n+1)\widetilde{A}_1}\Bigg[33(n+2)\widetilde{B_2}(2\widetilde{B_1} +33\widetilde{B_2}) +(n+1)(153\widetilde{B_2}+4\widetilde{C_1}-1089\widetilde{C_3})\widetilde{A_1} \Bigg]\\
\widetilde{\widetilde{C}}_2 &=&\frac{-1}{4(n+1)\widetilde{A}_1}\Bigg[2(n+2)\widetilde{B_2}(2\widetilde{B_1} +33\widetilde{B_2}) +(n+1)(19\widetilde{B_2}-4(\widetilde{C_2}+33\widetilde{C_3}))\widetilde{A_1} \Bigg]
\end{eqnarray}
\end{widetext}
In deriving the above formulae, the following equation is implicitly adopted, i.e. the value of $a_s$ at any scale $Q^*$ can be obtained from its value at the scale $Q$,
\begin{widetext}
\begin{eqnarray}\label{alphaexp}
a_{s}(Q^*) &=& a_{s}(Q)-\frac{1}{4} \beta_{0} \ln\left(\frac{Q^{*2}}{Q^2}\right) a^{2}_{s}(Q) +\frac{1}{4^2}\left[\beta^2_{0} \ln^2 \left(\frac{Q^{*2}}{Q^2}\right) -\beta_{1} \ln\left(\frac{Q^{*2}}{Q^2}\right) \right] a^{3}_{s}(Q) + \nonumber\\
&& \frac{1}{4^3}\left[-\beta^3_{0} \ln^3 \left(\frac{Q^{*2}}{Q^2}\right) +\frac{5}{2} \beta_{0}\beta_{1} \ln^2\left(\frac{Q^{*2}}{Q^2}\right) -\beta_{2} \ln\left(\frac{Q^{*2}}{Q^2}\right)\right] a^{4}_{s}(Q) +{\cal O}(a^{5}_{s}(Q)) .
\end{eqnarray}
\end{widetext}


\begin{thebibliography}{99}

\bibitem{selfconsistence} S.J. Brodsky, SLAC-PUB-6304 (1993); S.J. Brodsky and H.J. Lu, SLAC-PUB-6000, arXiv:9211308.

\bibitem{qedlimit} S.J. Brodsky and P. Huet, Phys.Lett. B{\bf 417}, 145-153 (1998).

\bibitem{blm} S.J. Brodsky, G.P. Lepage and P.B. Mackenzie, Phys.Rev. D{\bf 28}, 228(1983).

\bibitem{pmc} S.J. Brodsky and L.D. Giustino, arXiv: 1107.0338.

\bibitem{scale1} S.J. Brodsky and H.J. Lu, Phys.Rev. D{\bf 51}, 3652(1995).

\bibitem{scale2} G. Grunberg, Phys.Rev. D{\bf 46}, 2228(1992).

\bibitem{pms} P.M. Stevenson, Phys.Lett. B{\bf 100}, 61(1981); Phys.Rev. D{\bf 23}, 2916(1981); Nucl.Phys. B{\bf 203}, 472(1982); Nucl.Phys. B{\bf 231}, 65(1984).

\bibitem{bl1} H.J. Lu and S.J. Brodsky, Phys.Rev. D{\bf 48}, 3310(1993).

\bibitem{beta} O.V. Tarasov, A.A. Vladimirov and A. Yu Zharkov, Phys.Lett. B{\bf 93}, 429(1980); T. van Ritbergen, J.A.M. Vermaseren and S.A. Larin, Phys.Lett. B{\bf 400}, 379(1997); M. Czakon, Nucl.Phys. B{\bf 710}, 485(2005).

\bibitem{lambert} E. Gardi, M. Karliner and G. Grunberg, JHEP {\bf 9807}, 007(1998).

\bibitem{tH} G.'t Hooft, in {\it The Whys of Subnuclear Physics}, Proceedings of the International School of Subnuclear Physics, Erice, Italy, 1977, edited by A. Zichichi, Subnuclear Series Vol.15 (Plenum, New York, 1979), p.943.

\bibitem{kataev} A.V. Garkusha and A.L. Kataev, Phys.Lett. B{\bf 705}, 400(2011).

\bibitem{conLam1} W.A. Bardeen, A.J. Buras, D.W. Duke and T. Muta, Phys.Rev. D{\bf 18}, 3998(1978).

\bibitem{conLam2} W. Furmanski and R. Petronzio, Z.Phys. C{\bf 11}, 293(1982).

\bibitem{lams1} W.J. Marciano, Phys.Rev. D{\bf 29}, 580(1984).

\bibitem{lams2} L.F. Abbott, Phys.Rev.Lett. {\bf 44}, 1569(1980)

\bibitem{lams3} E. Monsay and C. Rosenzweig, Phys.Rev. D{\bf 23}, 1217(1981).

\bibitem{fourloopa} K.G. Chetyrkin, B.A. Kniehl and M. Steinhauser, Phys.Rev.Lett. {\bf 79}, 2184(1997).

\bibitem{Kataev} G. Grunberg and A.L. Kataev, Phys.Lett. B{\bf 279}, 352(1992).

\bibitem{Mikhailov1} S.V. Mikhailov, JHEP {\bf 0706}, 009(2007).

\bibitem{Mikhailov2} A.L. Kataev and S.V. Mikhailov, Teor.Mat.Fiz. {\bf 170}, 174-186 (2012); arXiv:1011.5248[hep-ph].

\bibitem{eff} G. Grunberg, Phys.Lett. B{\bf 95}, 70 (1980); B{\bf 110}, 501(1982); Phys.Rev. D{\bf 29}, 2315(1984); A. Dhar and V. Gupta, Phys.Rev. D{\bf 29}, 2822 (1984).

\bibitem{gml} M. Gell-Mann and F.E. Low, Phys.Rev. {\bf 95}, 1300 (1954).

\bibitem{Ralphas} P.A. Baikov, K.G. Chetyrkin and J.H. Kuhn, Phys.Rev.Lett.{\bf 101}, 012002(2008); arXiv:0906.2987[hep-ph]; K. Nakamura et al. (Particle Data Group), J.Phys. G{\bf 37}, 075021 (2010).

\bibitem{fourloopb} R.V. Harlander and M. Steinhauser, Comput.Phys. Commun. {\bf 153}, 244(2003).

\bibitem{kataevC} A.L. Kataev, Pisma Zh.Eksp.Teor.Fiz. {\bf 94}, 867(2012); arXiv:1108.2898[hep-ph].

\bibitem{Chetyrkin} P.A. Baikov, K.G. Chetyrkin, J.H. Kuhn and J. Rittenger, work presented by K. G. Chetyrkin at 10th International Symposuim RADCOR2011 on Radiative Corrections (Applications to Quantum Field Theory to Phenomenology), Mamallapuram, India, September 29, 2011.

\bibitem{rexp} R. Marshall, Z.Phys. C{\bf 43}, 595 (1989).

\bibitem{pdg} K. Nakamura, {\it et al.}, Particle Data Group, J.Phys. G{\bf 37}, 075021(2010).

\bibitem{cleo} R. Ammar etal. (CLEO Collaboration), Phys.Rev. D{\bf 57}, 1350(1998).

\bibitem{jet} G. Dissertori, {\it etal.}, JHEP {\bf 0802}, 040(2008).

\bibitem{aveas} S. Bethke, Eur.Phys.J. C{\bf 64}, 689 (2009).

\end{thebibliography}
\end{document}